\documentclass[prd,preprint,tightenlines,showpacs,preprintnumbers,nofootinbib,eqsecnum,superscriptaddress]{revtex4}

\usepackage[T1]{fontenc} 
\usepackage{amsfonts,amssymb,amscd,amsmath}
\usepackage[dvipsnames]{xcolor}
\usepackage{graphicx}
\usepackage{verbatim}

\newcommand{\rb}{\mbox{\boldmath $b$}}
\def\pom{{I\!\!P}}

\begin{document}


\title{Exclusive vector - toponium photoproduction in hadronic collisions}

\author{Victor P. {\sc Gon\c{c}alves}}
\email{barros@ufpel.edu.br}
\affiliation{Institute of Physics and Mathematics, Federal University of Pelotas, \\
  Postal Code 354,  96010-900, Pelotas, RS, Brazil}

\author{Luana {\sc Santana}}
\email{luanas1899@gmail.com}
\affiliation{Institute of Physics and Mathematics, Federal University of Pelotas, \\
  Postal Code 354,  96010-900, Pelotas, RS, Brazil}

\author{Bruno D. {\sc Moreira}}
\email{bduartesm@gmail.com }
\affiliation{Departamento de F\'isica, Universidade do Estado de Santa Catarina, 89219-710 Joinville, SC, Brazil.}

\date{\today}

\begin{abstract}
 An exploratory study of the  exclusive production of a vector - toponium $\psi_t$ state by photon - hadron interactions  is performed considering  proton - proton and proton - nucleus collisions at the Large Hadron Collider (LHC) and Future Circular Collider (FCC) energies. The scattering amplitude is calculated using the $k_T$ - factorization formalism assuming that the vector - toponium state can be described by a Gaussian light-cone wave function and  considering different models for the unintegrated gluon distribution.  Predictions for the rapidity distributions and total cross - sections are presented.  Our results indicate that the experimental measurement of this final state  will be very difficult for the expected integrated luminosities at the LHC and FCC. 

\end{abstract}

\maketitle


\section{Introduction}
One of the main challenges of strong interactions theory is the description of properties and the understanding of production mechanisms of heavy quark bound (quarkonium) states, as e.g. the $\eta_c$ and $J/\Psi$ particles, which are the lowest $S$ - wave  $c\bar{c}$ bound states characterized by distinct $J^{PC}$ quantum numbers \cite{Lansberg:2019adr}. 
Over the last decades, the theoretical treatment and the experimental investigation of the $c\bar{c}$ and $b\bar{b}$ bound states have been largely advanced (For a review see, e.g.   Ref. \cite{Brambilla:2010cs}).  In contrast,  signals of a toponium state, formed  by a top and an antitop quark, were only recently observed in $pp$ collisions at the LHC \cite{ATLAS:2019hau,CMS:2025kzt}, although its formation was proposed many years ago \cite{Fadin:1990wx,Fabiano:1993vx}. The observation of such events, which are consistent  with the formation of a pseudoscalar $t\bar{t}$ state ($\eta_t$ state), have motivated a vast phenomenology  \cite{Aguilar-Saavedra:2024mnm,Fuks:2024yjj,Garzelli:2024uhe,Wang:2024hzd,Fu:2024bki,Jiang:2024fyw,Francener:2025tor}.

A natural next question is if the observation of a  vector toponium state $\psi_t$ is feasible at the LHC and future hadronic colliders. 
Differently from the $\eta_t$ state, which can be directly produced   in hadronic collisions through the reaction $gg \rightarrow \eta_t$, such channel is prohibit in the $\psi_t$ case due to the Landau - Yang theorem \cite{Landau:1948kw,Yang:1950rg}. As a consequence, in order to investigate the  $\psi_t$ production by gluon - gluon interactions, we have to consider higher order processes, as e.g. those associated with the reactions $gg \rightarrow \psi_t g$ or $gg \rightarrow \psi_t b \bar{b}$. Such possibility was investigated in  Ref. \cite{Bai:2025buy}, where the authors have estimated the corresponding cross - sections in inclusive reactions, where both incident proton fragment, and concluded that the experimental separation of the associated events at both the current LHC and the future high - luminosity run (HL - LHC) will be a hard task.

\begin{figure}[t]
    \centering
    \includegraphics[width=0.65\linewidth]{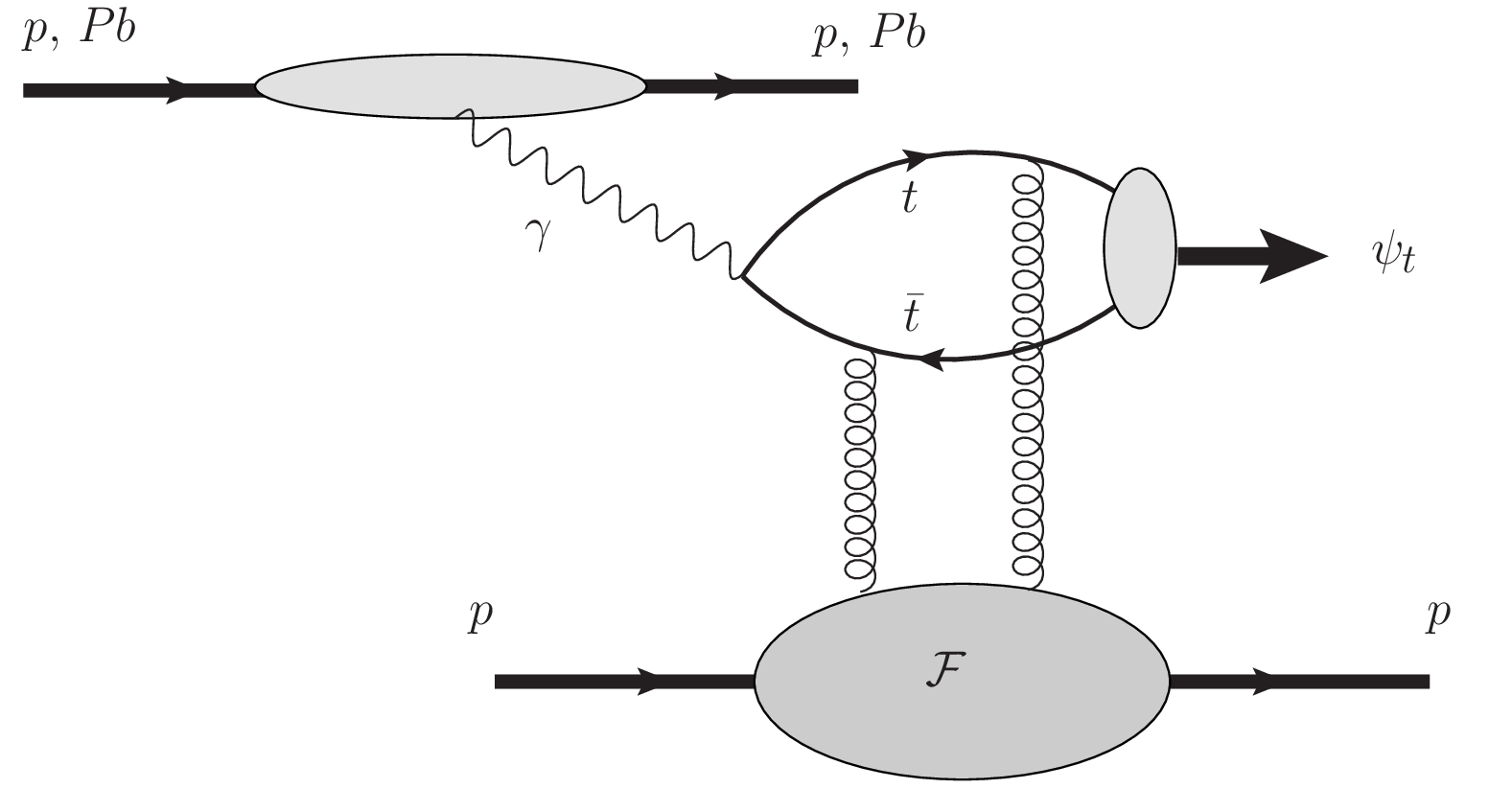}
    \caption{Exclusive vector - toponium photoproduction in  $pp$ and $pPb$ collisions within the $k_T$ - factorization formalism.}
    \label{fig:diagram}
\end{figure}

An alternative to investigate the quarkonium production in hadronic collisions is the study of photon - induced interactions, which became a reality in recent years \cite{upc}. 
In this paper, motivated by the experimental results for the exclusive $J/\Psi$ and $\Upsilon$ photoproduction in hadronic collisions,  we will perform { an exploratory study and investigate, for the first time,} the exclusive $\psi_t$ photoproduction in $pp$ and $pPb$ collisions at the LHC and FCC energies. Considering that both incident hadrons can act as photons sources and hadronic targets, the differential cross - section for the exclusive $\psi_t$ photoproduction in a collision between two hadrons ($h_1h_2$) will be given in the equivalent photon approximation (EPA) \cite{epa} by
\begin{eqnarray}
\frac{d\sigma \,\left[h_1 h_2 \rightarrow   h_1 +  \psi_t + h_2\right]}{d^2\rb \, dy_{\psi_t}} = \omega_{h_1} N_{h_1}(\omega_{h_1},b)\,\sigma_{\gamma h_2 \rightarrow \psi_t \otimes h_2}\left(\omega_{h_1} \right) + (1\longleftrightarrow2) \,\,,
\label{dsigdy}
\end{eqnarray}
where $\rb$ is the impact parameter of the collision ($b = |\rb|$), $y_{\psi_t}$ is the rapidity  of $\psi_t$ vector meson in the final state, and $\omega_{h_i}$  is the energy of the photon emitted by the hadron $h_i$. We have that $\omega_{h_1} = (M_{\psi_t}/2)e^{y_{\psi_t}}$ and $\omega_{h_2} = (M_{\psi_t}/2)e^{-y_{\psi_t}}$, with $M_{\psi_t}$ the mass of the vector toponium state, and $N(\omega,b)$ represents the photon spectrum.
Moreover, $\sigma_{\gamma h_i \rightarrow \psi_t \otimes h_i}$ is the cross - section for the production of a $\psi_t$ state in a photon - hadron interaction, and the symbol
$\otimes$ represents that this interaction was mediated by a color singlet object, usually denoted Pomeron $\pom$, and  that a rapidity gap is expected to be present in the final state. In our analysis, we will estimate the photoproduction cross - section using the $k_T$ - factorization formalism, proposed originally in Refs. \cite{Ivanov:1999pb,Ivanov:2004ax} and successfully applied for hadronic collisions, e.g., in Refs. \cite{Schafer:2007mm,Rybarska:2008pk,Cisek:2010jk,Cisek:2014ala,
Luszczak:2019vdc,Bolognino:2019pba}. In this formalism, the exclusive $\psi_t$ photoproduction in hadronic collisions is represented as in Fig. \ref{fig:diagram}. The main ingredients in the calculation are the unintegrated gluon distribution $\cal{F}$ and the wave function for the $\psi_t$ state. We will consider two different models for $\cal{F}$, based on distinct assumptions for the QCD dynamics, and we will derive the light cone wave function assuming a Gaussian ansatz and using the recent results for the radial wave function at the origin derived in Ref. \cite{Bai:2025buy}. 

The analysis performed in this paper is also motivated by the possibility of searching for events associated with the exclusive $\psi_t$ photoproduction using the existing forward proton detectors.
Over the last years,  exclusive events  in $pp$ collisions have been separated at the LHC by measuring the outgoing protons using the forward proton
detectors (FPD), such as the ATLAS Forward Proton detector (AFP)
\cite{Adamczyk:2015cjy,Tasevsky:2015xya} and Precision Proton Spectrometer (CT-PPS) \cite{Albrow:2014lrm}, that are installed symmetrically around the interaction point at a distance of roughly 210~m from the interaction point. Such analyzes were restricted to final states characterized by a large invariant mass ($M > 200$ GeV) due to the current limitations in the reconstruction of the protons in the final state. As a consequence,  the current forward proton detectors are not able to separate exclusive events associated with $J/\Psi$ and $\Upsilon$ photoproduction by photon - Pomeron interactions. However, such a limitation is not present in the case of a vector toponium state.


This paper is organized as follows. In the next Section, we will present a brief review of the $k_T$ - factorization formalism and the main ingredients used in our calculations. In particular, we will discuss the description of the unintegrated gluon distribution and derive the light cone wave function for the $\psi_t$ state. In Section \ref{sec:res} we will present our predictions for the total cross - sections and rapidity distributions, derived considering $pp$ and $pPb$ collisions at the LHC and FCC energies. Finally, in Section \ref{sec:sum}, we will  summarize our main results and conclusions.

\section{Formalism}
In what follows, we will discuss in more details the ingredients needed to calculate the differential cross - section for the exclusive $\psi_t$ photoproduction in $h_1 h_2$ collisions, as given in Eq. (\ref{dsigdy}). For the photon spectrum,
$N(\omega,b)$, we will consider that it can be expressed in terms  of the charge form factor $F(q)$ as follows \cite{upc}
\begin{eqnarray}
 N(\omega,b) = \frac{Z^{2}\alpha}{\pi^2}\frac{1}{b^{2} v^{2}\omega}
\cdot \left[
\int u^{2} J_{1}(u) 
F\left(
 \sqrt{\frac{\left( \frac{b\omega}{\gamma_L}\right)^{2} + u^{2}}{b^{2}}}
 \right )
\frac{1}{\left(\frac{b\omega}{\gamma_L}\right)^{2} + u^{2}} \mbox{d}u
\right]^{2} \,\,,
\label{Eq:fluxo0}
\end{eqnarray}
where $\alpha$ is the electromagnetic coupling constant, $\gamma_L$ is the Lorentz factor and $v$ is the hadron velocity.
For the nucleus, we will estimate the photon flux using the realistic form factor, which corresponds to the Wood - Saxon distribution and is the Fourier transform of the charge density of the nucleus. On the other hand, for the proton case, we will assume that $F$ is given by the dipole form factor. In addition, we will focus on ultraperipheral hadronic collisions and integrate Eq. (\ref{dsigdy}) over $\rb$ in the range $b  \ge b_{min}$, with $b_{min} = R_{h_1} + R_{h_2}$. 

The exclusive $\psi_t$ photoproduction cross - section in photon - proton interactions, $\sigma_{\gamma p \rightarrow \psi_t \otimes p}$, will be given by
\begin{eqnarray}
    \sigma_{\gamma p \rightarrow \psi_t \otimes p} = \int dt\frac{d\sigma}{dt} = \int dt \left.\frac{d\sigma} {dt}\right|_{t=0}e^{Bt} =  \frac{1}{16\pi B} \left|  {\mathrm{Im}\mathcal{M}(W,t = 0)} \right |^2 \,\,,
    \label{eq:proton}
\end{eqnarray}
where $B$ is the slope parameter and $W$ the photon - proton center - of - mass energy. Moreover, the imaginary part of the forward scattering amplitude,  ${\mathrm{Im}\mathcal{M}(W,t = 0)}$, is expressed in the $k_T$ - factorization formalism \cite{Ivanov:1999pb,Ivanov:2004ax} as follows
\begin{eqnarray}
    \mathrm{Im} \mathcal{M} & = & \frac{2\sqrt{4\pi\alpha_{em}}c_V}{4\pi^2} \int_0^1  \frac{\mathrm{d}z}{z(1-z)} \int_0^{\infty}\pi \mathrm{d}\boldsymbol{\kappa}^2\psi_V(z,\boldsymbol{\kappa}^2) 
    \int_0^{\infty} \frac{\pi\mathrm{d}\mathbf{k}_T^2}{\mathbf{k}_T^4}\alpha_s(\mathbf{k}_T^2){\cal{F}}_p(x_{eff},\mathbf{k}_T^2) \nonumber \\  &\times & \left [ A_0(z,\boldsymbol{\kappa}^2) W_0(\boldsymbol{\kappa}^2,\mathbf{k}_T^2)+A_1(z,\boldsymbol{\kappa}^2)W_1(\boldsymbol{\kappa}^2,\mathbf{k}_T^2) \right ],
    \label{im_M_kt}
\end{eqnarray}
where $c_V=2/3$, $z$ is the fraction of the photon light - cone momentum carried by the top quark and $\boldsymbol{\kappa}$  the corresponding transverse momentum. The function 
${\cal{F}}_p(x_{eff},\mathbf{k}_T^2)$ is the proton unintegrated gluon distribution (UGD), which depends only on the gluon longitudinal momentum fraction, $x_{eff}={M^2_{\psi_t}}/{W^2}$, and transverse momentum $\mathbf{k}_T$, and is determined by the QCD dynamics.
Moreover, the auxiliary functions $A_0$, $A_1$, $W_0$ and $W_1$ are defined by
\begin{eqnarray}
        A_0(z,\boldsymbol{\kappa}^2) &= & m_t^2+\frac{\boldsymbol{\kappa}^2m_t}{M+2m_t}, \\
    A_1(z,\boldsymbol{\kappa}^2) &= & \left [ z^2+(1-z)^2-(2z-1)^2\frac{m_t}{M+2m_t} \right ]\frac{\boldsymbol{\kappa}^2}{\boldsymbol{\kappa}^2+m_t^2}, \\
    W_0(\boldsymbol{\kappa}^2,\mathbf{k}_T^2) &=& \frac{1}{\boldsymbol{\kappa}^2+m_t^2}-\frac{1}{\sqrt{(\boldsymbol{\kappa}^2-m_t^2-\mathbf{k}_T^2)^2+4m_t^2\boldsymbol{\kappa}^2}}, \\
    W_1(\boldsymbol{\kappa}^2,\mathbf{k}_T^2) &=& 1-\frac{\boldsymbol{\kappa}^2+m_t^2}{2\boldsymbol{\kappa}^2}\left [ 1+\frac{\boldsymbol{\kappa}^2-m_t^2-\mathbf{k}_T^2}{\sqrt{(\boldsymbol{\kappa}^2-m_t^2-\mathbf{k}_T^2)^2+4m_t^2\boldsymbol{\kappa}^2}} \right ],
\end{eqnarray}
where $m_t$ is the top mass. For the case of photon - nucleus interactions, the total cross - section will be estimated assuming that 
\begin{eqnarray}
      \label{sigma_mesons}
    \sigma(\gamma + A \rightarrow V  \otimes A)=\frac{1}{16\pi }\left |  {\mathrm{Im}\mathcal{M}}({W^2,t=0)} \right |^2\int_{t_{min}}^{\infty}\mathrm{d}t|F_A(t)|^2 \,\,,
\end{eqnarray}
where $F_A(t)$ is nuclear form factor, and that the unintegrated gluon distribution for the nucleus can be expressed in terms of ${\cal{F}}_p$ as follows: ${\cal{F}}_A(x_{eff},\mathbf{k}_T^2) = A \times {\cal{F}}_p(x_{eff},\mathbf{k}_T^2)$. Therefore,  we will disregard, in a first approximation, the contribution of nuclear effects, which are expected to modify the behavior of the gluon distribution in a nuclear medium \cite{Armesto:2006ph}. In order to estimate the dependence of our predictions on the parameterization assumed for ${\cal{F}}_p$,  we will consider the parameterizations for ${\cal{F}}_p$ obtained in Refs. \cite{BermudezMartinez:2018fsv}  and   \cite{Hautmann:2013tba}     by fitting the HERA data assuming that the evolution is described by the DGLAP equation + parton branching method and CCFM equation, respectively. Both parameterizations are available in the   TMDlib website\footnote{https://tmdlib.hepforge.org/}, and will be denoted HERA I+II and CCFM hereafter.  

The last ingredient needed to estimate the scattering amplitude is the light - cone wave function of the vector toponium. As in Refs. \cite{Ivanov:2004ax,Cisek:2014ala}, we will assume a Gaussian ansatz:
\begin{eqnarray}
        \label{gaussiana}
    \psi(\vec{p\,}^2)= C\exp\left (-\frac{\vec{p\,}^2 a^2}{2} \right )\,\,,
\end{eqnarray}
where $\vec{p}$ is the relative momentum of $t$ and $\bar{t}$ in the rest frame of meson, which can expressed in terms of $z$ and $\boldsymbol{\kappa}$: $\vec{p} = (\mathbf{p},p_z) = (\boldsymbol{\kappa},(z-1/2)M_{\psi_t})$. The parameters $C$ and $a$ are usually obtained by fitting to the vector meson decay width into an $e^+e^-$ pair, expressed by 
\begin{eqnarray}
    \Gamma(\psi_t \rightarrow e^+e^-)=\frac{4\pi \alpha^2c_V^2} {3M_{\psi_t}^3}\cdot f_V^2 \,\,,
    \label{eq:gama}
\end{eqnarray}
where 
\begin{eqnarray}
        f_V \equiv \frac{N_c}{(2\pi)^3}\int \mathrm{d}^3\mathbf{p}\frac{8}{3}(M_{\psi_t}+m_t)\psi(\mathbf{p}^2) \,\,,
\end{eqnarray}
and by imposing the normalization condition:
\begin{eqnarray}
      \frac{N_c}{(2\pi)^3}\int \mathrm{d}^3\mathbf{p}\,4M_{\psi_t}|\psi(\mathbf{p}^2)|^2 = 1 \,\,.
      \label{eq:norm}
\end{eqnarray}
As the electronic decay width for the $\psi_t$ state has not yet been measured, we will estimate its value using that it can be expressed in terms of the radial wave function at the origin, $R_S(0)$, as follows~\cite{Eichten:1995ch}
\begin{eqnarray}
      \Gamma(\psi_t\rightarrow e^+e^-)=\frac{4N_c\alpha^2c_V^2}{3}\frac{|R_S(0)|^2}{M_{\psi_t}^2}\left(1-\frac{16\alpha_s} {3\pi}\right) \,\,.
\end{eqnarray}
In our analysis, we will assume the value for $R_S(0)$ derived in Ref. \cite{Bai:2025buy} by solving the non - relativistic Schr\"odinger equation assuming a static potential. It implies $\Gamma(\psi_t\rightarrow e^+e^-)= 9.962$ keV, which is similar to the values obtained in Refs. \cite{Yndurain:2000yq,Wang:2024hzd}. Using this value and the conditions given in Eqs. (\ref{eq:gama}) and (\ref{eq:norm}) we can derive the values of $C$ and $a$ for our ansatz for the light - cone wave function of the $\psi_t$ state. It is important to emphasize that the analysis performed in Ref. \cite{Cisek:2014ala} has demonstrated that the Gaussian ansatz for the $J/\psi$ meson provides a quite good description of the HERA and LHC data.

\section{Results}
\label{sec:res}
In what follows, we will present our predictions for the exclusive $\psi_t$ photoproduction in $pp$ and $pPb$ collisions at the LHC and FCC energies. In our calculations, as in Ref. \cite{Bai:2025buy},   we will assume 
$m_t=172.4$ GeV and $M_{\psi_t}=342.1$ GeV. The value of the slope parameter $B$ for the exclusive $\psi_t$ photoproduction is an open question. The experimental results from HERA indicate that this value decreases for heavier quarkonium states \cite{Ivanov:2004ax}.  Considering this trending, we  expect that $B \le 4.0$ GeV$^{-2}$. As this parameter determines the normalization of the cross - section, we will consider, for simplicity, that $B = 1$ GeV$^{-2}$ in Eq. (\ref{eq:proton}), which easily allows the rescaling of our predictions if other value is assumed. 
{ However, it is important to emphasize that such assumption for the value of $B$ implies that our predictions for the magnitude of the cross - section are optimistic and must be considered as an upper bound.   }

\begin{figure}[t]
    \centering
    \includegraphics[width=0.65\linewidth]{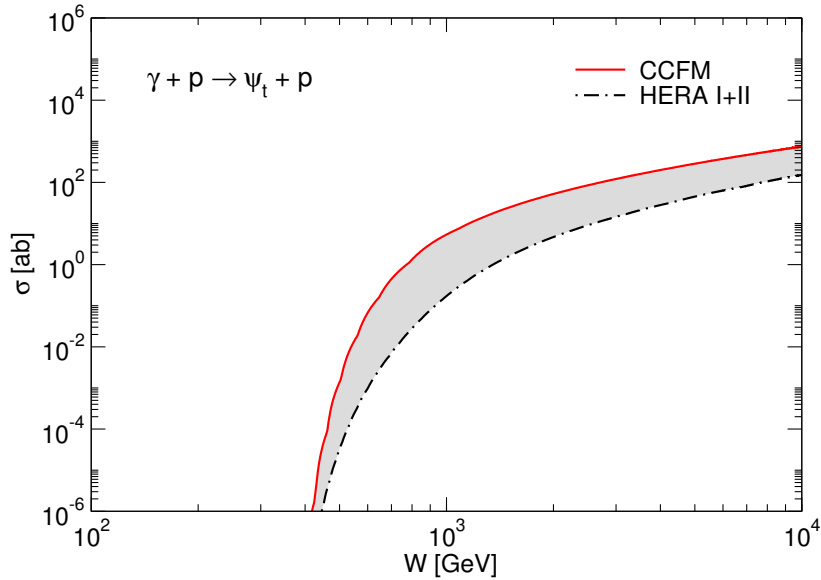}
    \caption{Predictions for the energy dependence of the exclusive $\psi_t$ photoproduction cross - section in photon - proton interactions, derived considering two different parameterizations for the proton unintegrated gluon distribution.}
    \label{fig:photonproton}
\end{figure}

In Fig. \ref{fig:photonproton} we present our predictions for the dependence of the exclusive $\psi_t$ photoproduction cross - section  on the photon - proton center - of - mass energy $W$, derived considering the HERAI+II and CCFM parameterizations for the unintegrated gluon distribution. As expected from the results for the production of lighter quarkonium states, the cross - section increases with the energy, with the slope being dependent on the UGD considered. We have that the HERAI+II parameterization implies smaller values in comparison with the CCFM one. In particular, for the kinematical range that can be probed in the proposed Large Hadron - electron Collider (LHeC) \cite{LHeC:2020van} and FCC - eh \cite{FCC:2018vvp,FCC:2025lpp}, we predict that the cross - section will be of the order of few ab. Considering the huge integrated luminosity expected to be achieved in these future colliders,  such result motivates a dedicated analysis of the feasibility of measuring the $\psi_t$ state in $\gamma p$ interactions at the LHeC and FCC - eh, which we plan to perform in a forthcoming study.

\begin{figure}[t]
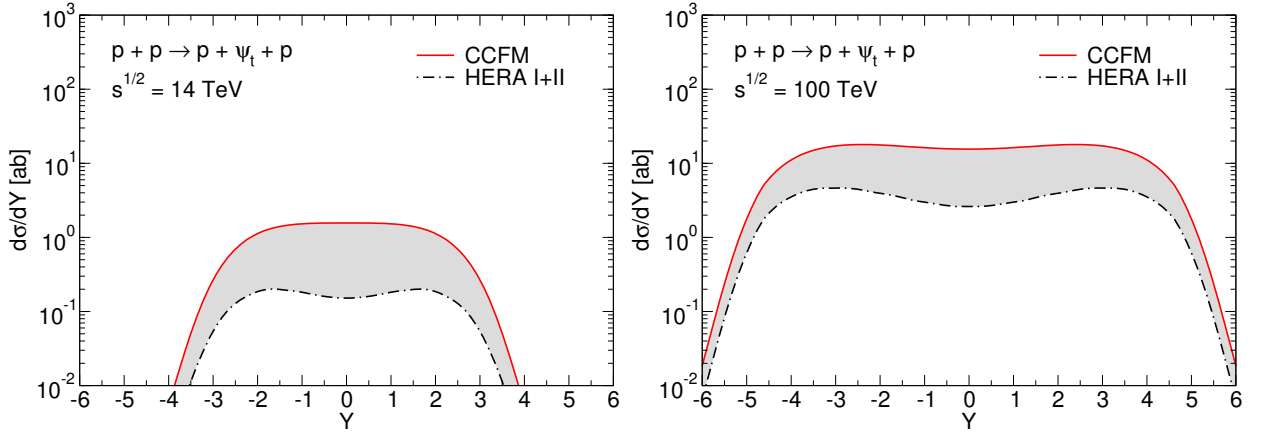

    \centering
    \includegraphics[width=0.49\linewidth]{toponium_pp_14TeV.eps}
    \includegraphics[width=0.49\linewidth]{toponium_pp_100TeV.eps}
    \caption{Predictions for the rapidity distribution associated with the exclusive $\psi_t$ photoproduction in $pp$ collisions at the LHC (left panel) and FCC (right panel) energies. Predictions derived considering different parameterizations for the proton UGD.}
    \label{fig:rap_pp}
\end{figure}

\begin{figure}[t]
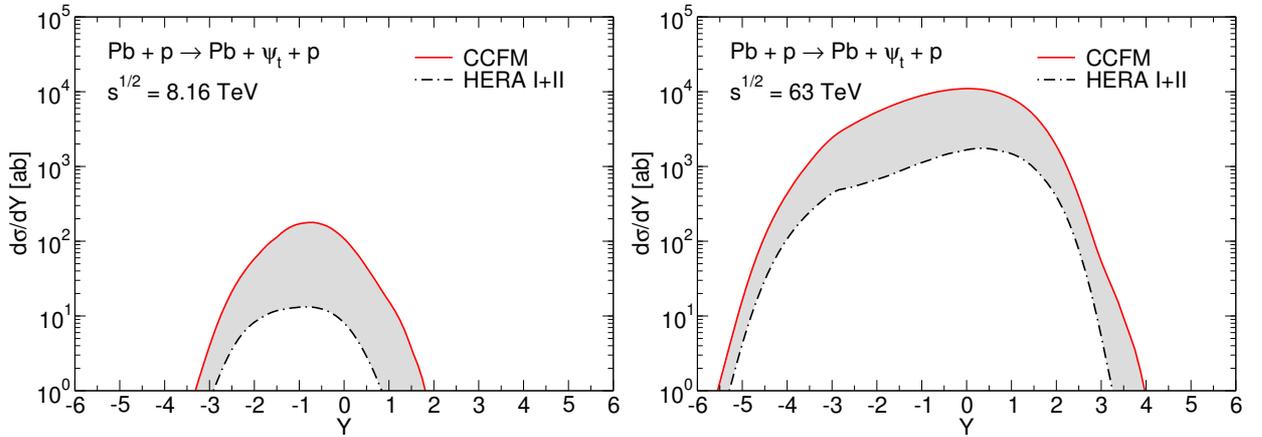

    \centering
    \includegraphics[width=0.49\linewidth]{toponium_Pbp_8_16TeV.eps}
    \includegraphics[width=0.49\linewidth]{toponium_Pbp_63TeV.eps}
    \caption{Predictions for the rapidity distribution associated with the exclusive $\psi_t$ photoproduction in $pPb$ collisions at the LHC (left panel) and FCC (right panel) energies. Predictions derived considering different parameterizations for the proton and nuclear UGD's.}
    \label{fig:rap_pPb}
\end{figure}

In Figs. \ref{fig:rap_pp} and \ref{fig:rap_pPb} we present our predictions for the rapidity distribution associated with the exclusive $\psi_t$ photoproduction in $pp$ and $pPb$ collisions, respectively. The results for  LHC (FCC) energies are presented in the left (right) panels. The distributions become wider and have a larger normalization when the  nucleon - nucleon center - of - mass energy is increased. In the case of $pPb$ collisions, we predict asymmetric rapidity distributions. Such behavior is associated with the fact that, for a fixed rapidity, the photon fluxes for the proton and lead will be determined by different photon energies ($\omega_i$) and the cross - sections ($\sigma_{\gamma h_i}$) by distinct photon - hadron center - of - mass energies ($W^2_{\gamma h_i} = 2 \omega_i \sqrt{s_{h_1h_2}}$). In $\gamma p$ interactions the photon comes from the lead, with the photon flux being proportional to $Z^2$ ($Z = 82$), and the photoproduction cross section depends on the square of ${\cal{F}}_p$. On the other hand, in $\gamma Pb$ interactions, the photon comes from the proton and the photoproduction cross section is determined by the square of ${\cal{F}}_{Pb}$. As a consequence, asymmetric rapidity distributions are expected in $pPb$ collisions.



\begin{table}[t]
    \centering
    \begin{tabular}{|l|c|c|c|c|}
    \hline
    \hline
         & $\sqrt{s_{pp}}=14$ TeV  & $\sqrt{s_{pp}}=100$ TeV & $\sqrt{s_{Pbp}}=8.16$ TeV  & $\sqrt{s_{Pbp}}=63$ TeV \\ \hline
         \hline 
        CCFM & $6.789$ ($0.7756$) & $83.40$ ($36.77$) & $349.8$ ($4.051\times 10^{-2}$) & $35111.4$ ($607.6$) \\
        \hline 
        HERA I+II & $0.8712$ ($0.1421$) & $16.55$ ($10.14$) & $29.63$ ($6.357\times 10^{-4}$) & $5314.9$  ($120.9$) \\ 
    \hline
    \end{tabular}
    \caption{Predictions for the total cross - sections (in ab) associated with the exclusive $\psi_t$ photoproduction in $pp$ and $pPb$ collisions at the LHC and FCC energies, derived considering the typical rapidity range covered by a central (forward) detector.}
    \label{tab:total}
\end{table}

Finally, in Table \ref{tab:total}, we present our predictions for the total cross - sections (in ab) associated with the exclusive $\psi_t$ photoproduction in $pp$ and $pPb$ collisions at the LHC and FCC energies. We consider the typical rapidity ranges covered by  central ($-2.5 \le Y \le +2.5$) and forward ($2.0 \le Y \le 4.5$) detectors. For  $pp$ collisions at the LHC energy, we predict cross - sections of the order of few ab, with the HERAI+II providing a lower bound and being smaller if the rapidity range covered by a forward detector is considered. Such values become of the order of dozen of ab for FCC energies. On the other hand, for $pPb$ collisions at LHC (FCC), we predict an increasing of the cross - section by one (two) order of magnitude in comparison with the $pp$ results.

{ It is important to emphasize that the experimental separation of the events associated with the exclusive quarkonium photoproduction is, in general, performed using the dileptons generated in the quarkonium decay. Considering the expected luminosities  in the HL-LHC run and in the future hadronic colliders and the results derived in Ref.~\citep{Bai:2025buy}   for the branching fraction of the vector toponium into dileptons [${\cal{O}}(10^{-5})$], the predicted number of events at the LHC and FCC will be strongly suppressed, implying that the experimental measurement of this final state will be, in principle, a very hard task. Currently, we are investigating alternatives to separate these events and intend to present the results in a forthcoming study.     }

\section{Summary}
\label{sec:sum}
Although the formation of a toponium state in $e^-e^+$ collisions had been predicted many years ago, the discovery of the top quark with a large mass and rapid decay have reduced the interest about the associated bound state during the last three decades. The situation has recently changed with the observation in $pp$ collisions at the LHC of a substantial excess of events near the kinematical $t\bar{t}$ threshold, which are consistent with the formation of a pseudoscalar $\eta_t$ toponium state. 
Such events have been observed in inclusive processes, where both incident protons fragment in the collision. Considering the large background present in these processes, in Ref. \cite{Francener:2025tor} the authors have proposed the analysis of the $\eta_t$ production in exclusive hadronic interactions, where the incident hadrons remain intact  and two rapidity gaps are present in the final state, providing a cleaner environment to separate the toponium state. In this paper, we expand the analysis of exclusive processes as a way to searching for toponium and perform { an exploratory study} of the exclusive photoproduction  of a vector toponium $\psi_t$ state. Differently from the $\eta_t$ state, which can be exclusively produced by photon - photon and gluon - gluon interactions, the vector state  can only be produced by photon - Pomeron  interactions. In this paper, we have considered the $k_T$ - factorization formalism to describe the exclusive $\psi_t$ photoproduction and presented predictions for $pp$ and $pPb$ collisions at the LHC and FCC energies, derived assuming a Gaussian ansatz for the light - cone wave function of meson and considering different parameterizations for the unintegrated gluon distribution. In particular, we presented results for the rapidity distributions and estimates for the total cross - sections considering the typical rapidity ranges covered by central and forward detectors. Our results indicated that the cross - sections are of the order of few ab at the LHC, increasing to some dozen of ab at the FCC. { Considering the expected luminosities at the HL - LHC run and at the FCC and the quarkonium decay into dileptons, our results indicated that a future experimental separation of this final state will be, in principle, a very hard task. Such result motivates a more detailed analysis of the process, considering new strategies to separate the associated events as well as taking into account of e.g. realistic experimental cuts.
Such a study is currently being performed, and the results are expected to be released soon.}

\section*{Acknowledgments}
L.S.  and V.P.G. were  partially supported by CNPq, CAPES (Finance code 001), FAPERGS and  INCT-FNA (Process No. 464898/2014-5). B.D.M. was partially supported by FAPESC.

\end{document}